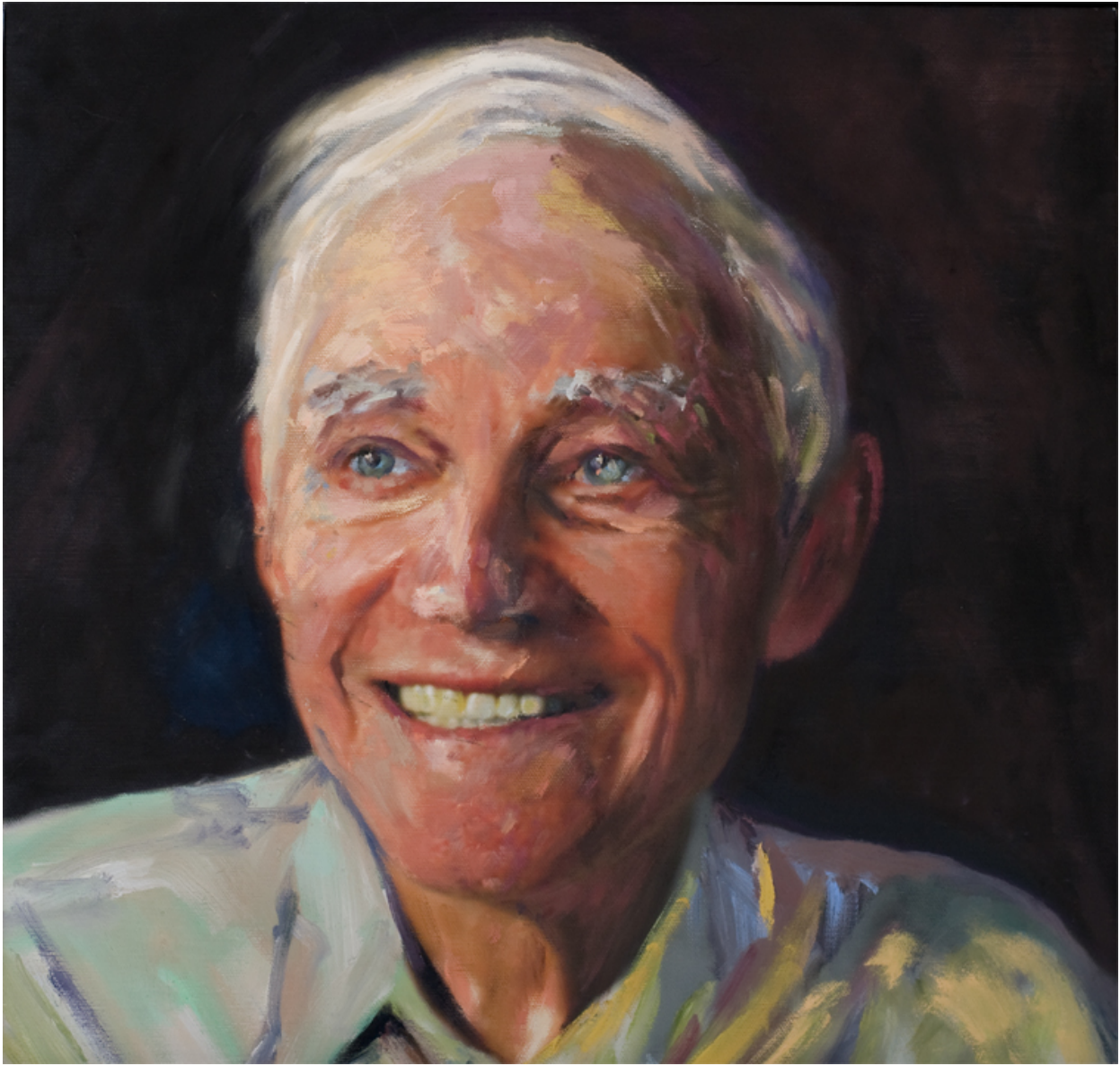

# Physics of Superconducting Transition Temperatures

Steven A. Kivelson

The superconducting transition temperature, $T_c$, is a dimensionful, non-universal quantity that manifestly depends on microscopic details. It depends on ``chemical'' differences between Hg and Pb, not just on whether the material in question is conventional or unconventional and whether or not there is a nearby quantum critical point or a fractionalized phase. For physicists interested in the theory of *superconductivity* or of *emergent quantum phenomena* in highly correlated electronic systems, $T_c$ may appear to be a ``microscopic detail'' that is unworthy of serious consideration. Indeed, since most conceivable superconducting phases have been classified, one might argue that superconductivity research should be transferred to Material Science and/or Chemistry Departments, and only studied in Physics Departments to the extent that superconductors are useful for realizing qubits and other elements of quantum circuits.

However, my friend, Ted Geballe, has always been interested first and foremost in $T_c$. I like talking to Ted, so I have been forced to think about $T_c$ as well. Ted will ask me "Why do you think cuprates with Hg in the charge reservoir layer have higher $T_c$'s?" or even more often will pose questions about promising but uncorroborated reports of USOs[1]. When I stop by his office to discuss what I learned at one meeting or another, he will ask me whether there was anything new about a second superconducting dome in very overdoped cuprates or whether everyone was discussing high temperature superconductivity in hydrides under pressure. (Usually, the answer I give is along the lines of, "No, mostly we discussed methods for braiding of non-Abelian anyons.")

In honor of my friend's 100$^{th}$ birthday, and in full recognition of the fact that ultimately the physics of $T_c$ cannot truly be addressed with the sort of statistical-mechanics-based approach to condensed matter physics I generally favor, I have put together some (possibly useful) thoughts concerning the physics of superconducting transition temperatures.

## I. Phase ordering vs pairing dominated $T_c$

As is well known, superconductivity requires both pairing and condensation. In conventional superconductors, $T_c$ is essentially the temperature at which pairs first form, but there are circumstances – for instance in granular superconductors or Josephson junction arrays – in which the pairing scale can be parametrically larger than $T_c$. In such systems, local (e.g. spectroscopic) measures of the material may still show what looks like a (possibly rounded) superconducting transition at a relatively high temperature, $T_{pair}$, but long-range phase coherence onsets only at $T_c < T_{pair}$.[1] In simple model problems, such as the negative U Hubbard model, it is possible to increase $T_{pair}$ without bound by simply increasing the strength of the pairing interaction. However, this inevitably leads to a crossover to a regime in which phase ordering – rather than pairing – determines $T_c$. In models, optimal $T_c$ always occurs as a crossover between a pairing dominated and a phase ordering dominated $T_c$.[2]

---

[1] Unidentified Superconducting Objects.

It is less clear how universally this observation applies in real materials. At a microscopic level, the dominant interaction between electrons is surely repulsive, and it is only through rather subtle collective effects that they can be tricked into pairing. Here is where retardation plays a central role in the theory of conventional superconductivity in allowing the relatively weaker electron-phonon mediated attraction to successfully out-compete the bare repulsion. However, there is some evidence that phase ordering plays an unusually large role in determining transition temperature of various "unconventional" superconductors with relatively high $T_c$'s, and even that optimal $T_c$ occurs as a crossover between a paring dominated and a phase ordering regime of parameters.[3]

## II. Negative U centers and superconductivity from preformed pairs

It is possible to imagine circumstances where a combination of good screening and/or strong electron-lattice coupling produce a strong net attraction. If the attractive interactions are strong enough, they can lead to relatively tightly bound states of pairs of electrons - and thus a low energy effective theory consisting of a fluid of bosonic "real-space pairs" or bipolarons. Naively, it would appear that looking for circumstances in which such pairs arise might provide fertile hunting grounds for high temperature superconductors.

If such attractions occur at special sites in a material – such as in the neighborhood of an impurity or an O vacancy – this can lead to negative U centers. As Ted and I recently reviewed[4], this notion was introduced by Anderson[5] to explain the properties of certain classes of disordered insulators, especially chalcogenide glasses. Because they are disordered and insulators, the electronic states tend anyway to be localized, so binding of a pair of electrons is substantially easier than in metals. The strong pairing explains a variety striking features of these materials, including the lack of Curie paramagnetism. However, the formation of pairs at negative U centers makes these materials more strongly insulating rather than superconducting!

For real-space pairing to occur in a translationally invariant system requires that the net attraction is stronger than the kinetic energy. If this arises from some form of strong electron-phonon coupling, it generally leads to very heavy pairs. Specifically, the effective hopping matrix element of a bipolaron is generically suppressed by a Frank-Condon factor,[6]

$$t^{eff} \sim (t^2/|U|) \exp[-\alpha |U|/\omega_0] \tag{1}$$

where t is the bare electron hopping matrix element, $|U|$ is the bipolaron binding energy, $\omega_0$ is an average of the phonon frequencies (measured in units of energy) and $\alpha \sim 1$. Since a prerequisite for real-space pairing is that $|U| > E_F$ (the bare Fermi energy) and since in most circumstance, $E_F \gg \omega_0$, bipolarons (and the pairs found at negative U centers) are effectively classical, non-dynamical objects – condensation of such pairs is implausible. This is why the chalcogenide glasses are particularly good insulators, not incipient superconductors.

If real-space pairing is to play any role in superconductivity, it must be in circumstances in which $\omega_0 \geq |U| \geq E_F$. Possibly, this can occur if the pairing is directly a consequence of electron correlations (no phonons) or if $E_F$ is unusually small (low electron density) and $\omega_0$ is relatively large. Otherwise, real-space pairing will always lead either to a superconducting state in which $T_c$ is parametrically small (due to low phase stiffness) or, more usually, to some competing insulating phase – either due to disorder or charge-density-wave (CDW) formation.[7]

### III. Upper bound on conventional superconductivity

Our basic understanding of conventional phonon-mediated superconductivity rests on sound theoretical principles.[9] These can be recast in the modern language of the renormalization group (RG), so long as all the underlying microscopic interactions are sufficiently weak (and short-ranged) – i.e. so long as a perturbative RG treatment is valid.[10,11] At a first stage, microscopic degrees of freedom with energies greater than the maximum phonon frequency, $\omega_{max}$, are integrated out. In this first stage of renormalization, the electron-phonon coupling is essentially unrenormalized, but the electron-electron repulsion renormalizes down from a bare value, $U$, to a renormalized value, $U^* = U/\{ 1 + \rho(E_F) U \log[ E_F/\omega_{max} ] \}$. At this stage, one is left only with electronic states in a narrow shell about the Fermi surface of width $\Delta k = \omega_{max}/v_F$ which interact with a renormalized interaction, $V_{eff} = U^* - U_{el-ph}$, in the "Cooper channel," where $U_{el-ph}$ is the induced attraction mediated by the exchange of phonons. In the second stage of RG, $V_{eff}$ is seen to be irrelevant (the ground-state is metallic) if $V_{eff} > 0$ and relevant (the ground-state is superconducting) if $V_{eff} < 0$, with $\lambda = -V_{eff} \rho(E_F)$ and $T_c \sim \omega_{max} \exp[ -1/\lambda ]$.

The success of this conventional understanding relies explicitly on the existence of a small parameter, $\omega_{max}/E_F$. Fortunately, this parameter is, in fact, typically very small in all conventional superconductors as it is proportional to one of the genuine small parameters nature provides us with, $\omega_{max}/E_F \sim [ m/M ]^{1/2}$, where m is the electron mass and M is the ion mass. However, it is often not recognized that in order for the perturbative RG to be valid, it is also necessary that $V_{eff} \ll E_F$, or in other words when $|\lambda| \ll 1$. So-called "strong-coupling" expressions for $T_c$ have been derived[12] that invoke Migdal's theorem as justification even when $\lambda > 1$ so long as $\lambda \ll E_F/\omega_{max}$. In all these expressions, $T_c$ is a monotonically increasing function of $\lambda$.

The electron-phonon problem is amenable to numerical solution by minus sign free determinantal quantum Monte Carlo. We have recently undertaken[13] an extensive such study of the simplest such model – the Holstein model – in which we can compare the exact results with those computed using "Migdal's approximation" – that is the approximate resumation of perturbation theory suggested by Migdal's theorem. We found that the Migdal approximation is remarkably accurate so long as $\lambda$ is smaller than a temperature dependent crossover value $\lambda^*(T) \sim \frac{1}{2}$, but that it breaks down both qualitatively and quantitatively for larger values of $\lambda$. This breakdown has to do with bipolaron formation and, for the reasons already discussed, this leads to an exponential suppression of any phase ordering tendency of the system. As a consequence, there is an optimal value of $\lambda$ – not too small (where the pairing scale is miniscule) and not too large (where phase coherence is quenched) – at which $T_c$ is maximal. This means that $T_c$ has a maximum magnitude (for given $\omega_{max}$) of the form[8]

$$T_c \leq \omega_{max} A \qquad (2)$$

Where A is a number of order 1. For the 2D Holstein model, we found empirically that A = 1/10.

With the caveat that there could well be special circumstances in which A is somewhat higher than this value, we proposed that this should be taken as an approximate bound for real materials with roughly the same value of A we obtained from the Holstein model. We have done an extensive (but possibly not exhaustive) search of published data on superconductors that are thought to be "conventional." Where available, we have taken the measured Debye temperature as a proxy for $\omega_{max}$, and otherwise we have taken $\omega_{max}$ from DFT calculations. In all cases we have found that this bound is satisfied – sometimes by a wide margin (in which case it is true but irrelevant) – while in other cases $T_c$ comes within 10%-20% of saturating this bound. Examples of materials that come close to this bound are Hg, Pb, $Nb_3Sn$, BKBO, and $H_3S$ under high pressure.

Manifestly, this bound is approximate. But the analysis that leads to it is sufficiently robust that it should be taken seriously. It suggests – not surprisingly – that to find high $T_c$ one needs to have the effective interactions that lead to pairing operating on a relatively high energy scale. Moreover, in common with all no-go results, its main usefulness may be to stimulate searches for ways to circumvent its reach.

## IV. Engineered composite systems

Another notion that may be useful in the search for new high temperature superconductors is the notion of using engineered composite systems as a platform. We have already seen that there is a trade-off as a function of strength of interactions between the imperative of having a high pairing scale and the need to retain a high phase coherence scale. One way around this is to devise a two-component system in which one portion of the system is strongly correlated, and is thus able to produce strong pairing between electrons, and the other is highly itinerant, and hence able to support a high coherence scale.[14]

Of course, Murphy's law implies that under most circumstances, such a composite system is likely to inherit the worst features of each subsystem. However, both theoretical and preliminary experimental evidence exists that – under appropriate circumstances – $T_c$ can be enhanced in such composite systems. (For a relatively recent review, see Ref. [15], and for some promising but not yet definitive experimental results suggesting that this strategy is possible, see [16-18].)

# V. Concerning some of Ted's favorite superconductors

Until now, I have dealt with generalities. Ted is a materials person. I have had endless discussions with him about how these various ideas apply in the context of specific materials.

### A. Negative U centers in Tl doped PbTe

Tl is a so-called "valence skipping" element. To the extent that a specific charge state can be identified in a solid, it generally is found either in a $Tl^{1+}$ or a $Tl^{3+}$ state. A many-body definition of the effective interaction between two electrons on a Tl site is given by

$$U = E(Tl^{1+}) + E(Tl^{3+}) - 2\, E(Tl^{2+}) \qquad (3)$$

where $E(Tl^{p+})$ is the energy of a Tl with charge +pe. Needless to say, for a Tl atom in free space, U defined in this way is positive; however, due to correlation effects associated with filled shells, it is anomalously small.[19] One can imagine that if the long-range part of the Coulomb interaction were screened, as it is in a solid, that this could give rise to an effective negative U.

Ted combined this insight with the observation that Tl doped PbTe was known to exhibit a superconducting $T_c$ that is considerably larger than any comparably low-density doped semiconductor, and concluded that the mechanism must be identified with Tl negative U centers.[20] In collaboration with Ian Fisher, they explored this idea, and have amassed considerable evidence in favor of this view.[21-23] In particular, evidence of a so-called "charge-Kondo effect" in the normal state fairly directly establishes the existence of negative U centers. Moreover, the superconductivity onsets at a critical value of the doping which can be seen to correspond to the point at which the Fermi energy satisfies the resonance condition, $2E_F = E(Tl^{1+}) + E(Tl^{3+})$, at which point Tl valence fluctuations between these two charge states should be observable.[23]

What is meant by a negative U center in this context? At minimum, what this means is that the effective U in Eq. (3) is negative and comparable to or larger in magnitude than $E_F$. In turn, U generally receives contributions from electron-correlation effects[27], but also from coupling to relatively high frequency phonons. However, the restriction to high-frequency phonons is essential. Otherwise the pair tunneling rate, $t_{eff}$, which in the charge-Kondo problem plays the role of the Kondo coupling,[24-26] is reduced by an exponential factor as in Eq. (1) – with $|U|$ replaced by $|U_{el-ph}|$, the electron-phonon contribution to U. Thus, given that the ions involved are relatively heavy, in order that -U is at least comparable to $E_F$ it is probably necessary that the dominant contribution to U is from some form of electronic overscreening. A negative U from strong electron-phonon coupling would tend to reproduce the physics of the chalcogenide glasses.

There is not seen, nor is there expected any large deviations from the BCS predictions for the superconducting state itself, since the superconducting correlation length is always large compared to the distance between Tl dopants. This means that for this purpose, the negative U centers can be viewed as giving rise to a uniform effective attraction of order x $t_{eff}$, where x is the concentration of Tl dopants. However, the mechanism involved – assuming all the inferences drawn by Geballe and Fisher hold up – is extremely unconventional. Although the superconducting order parameter is s-wave, the pairing is primarily derived from the electron-electron repulsion, and is not critically dependent on the electron-phonon coupling. This result should be more widely discussed.

### B. O vacancies in STO

Lightly doped $SrTiO_3$ is another of Ted's favorite superconductors.[28] It is certainly unusual; the very low carrier density at which superconductivity arises is more or less unique and the fact that the phonon frequencies are typically greater than the Fermi energy means that – even were the pairing mechanism phonon mediated – the way it plays out must be entirely different from that in conventional metallic superconductors.

While many people have focused on this and related aspects of this problem, Ted has long been excited by a possibly unrelated observation: The superconducting regime in STO doped with O vacancies extends to significantly lower doping concentrations than with the more usual Nd doping. Since most theoretical approaches to the problem involve no particular role of the dopant – other than to provide the requisite doped holes – understanding this manifestly requires new considerations. One could imagine that Nb impurities and O vacancies could disturb the electronic structure (introduce disorder) to different degrees, but there is no evidence of this apparent from normal state properties.

So, Ted focused on the structure of the impurities themselves and based on quantum chemical intuition suggested that the O vacancies may act as negative U centers, thus contributing to the pairing tendencies at very low doping. As with Tl, presumably what this means is that the local electronic structure leads to a significant reduction of the short-range repulsion between two electrons on a vacancy site, and that this reduced repulsion is converted to a net attraction due to environmental screening. Moreover, so long as the majority of this environmental screening is electronic in origin, or involves relatively small amplitude motions of optical phonon modes, it might not lead to a large Frank-Condon suppression of coherence. There are a number of ways to test this proposal in analogy to studies that have been carried out in Tl doped PbTe – this would be well worthwhile.

### C. Hydrides under pressure and a second dome in the cuprates?

There are many other anomalous superconductors that have interested Ted over the years. Naturally, he has been excited from the very first reports of high temperature superconductivity in hydrides at high pressure.[29-32] I am not aware of any new insights Ted has had here, but on the other hand one of the key players in this field was his grandson, Zach Geballe. So indirectly Ted has played a key role.

In contrast, Ted was one of the first to realize that there were a number of reports that there is a second superconducting dome in the cuprates at much higher hole doping than was previously appreciated.[33] One of the most significant observations about superconductivity in the cuprates is that it appears in some way linked to the antiferromagnetic order that characterizes the undoped parent compounds. This linkage has been corroborated in several ways, but the most intuitively clear was the observation that the superconducting dome always ends at an upper critical doping of somewhat less than x=0.3. It has been widely understood that beyond this point, the memory of the antiferromagnetic order in the parent compound is sufficiently weak that it can no longer underlie a mechanism of pairing. The observation[34-37] of superconductivity – with transition temperatures less than but of order 100K – for doping concentrations estimated to be x ~ 0.5 thus might indicate that there is something missing in the zeroth order understanding of the origin of superconducting pairing in the cuprates.

This is clearly the problem that most interests Ted at present. All of the materials involved are messy in some way or other – mostly because they involve complicated high-pressure synthesis that makes obtaining substantial amounts of material, much less single crystals, very difficult. There is clearly much work that needs to be done to characterize these materials, their normal state electronic structure, and the nature of the superconducting state that arises. But Ted is clearly right that this is an important direction for future research.

**VI. Ted Geballe**

The two portraits of Ted Geballe reproduced in this article are oil paintings on canvas by Pamela Davis Kivelson. (Her legal name is Pamela Davis, but her art name is Pamela Davis Kivelson, or PDK for short.) The first is a formal portrait that oversees the common area of the Geballe Laboratory for Advanced Materials (GLAM) at Stanford. Ted posed patiently for this portrait while my daughter – Sophia (then probably about 10 years old) - entertained him with songs from Gilbert and Sullivan operettas. It accurately captures something of Ted's personality. It is a smiling, happy portrait which is certainly rare in the history of art; it makes one joyful just to see it, much the way everyone who knows Ted is uplifted by each discussion with him. But I like the second portrait even better. It is Ted from behind – and all that can be seen is his familiar hat and red-checked jacket. I do not know whether it is the angle of the head or the slope of the shoulders, or just the familiarity of the simple items of clothing – at any rate, there is no mistaking that this is Ted.

Pam is a psychological painter – she cannot help herself. These two pictures unmistakably reflect her unadulterated love of the subject. But then, everyone loves Ted.

Acknowledgements: I benefitted from helpful comments from I. R. Fisher and K. Behnia. This synergistic work was supported in part by the U. S. Department of Energy (DOE) Office of Basic Energy Science, at Stanford under contract No. DE-AC02-76SF00515.

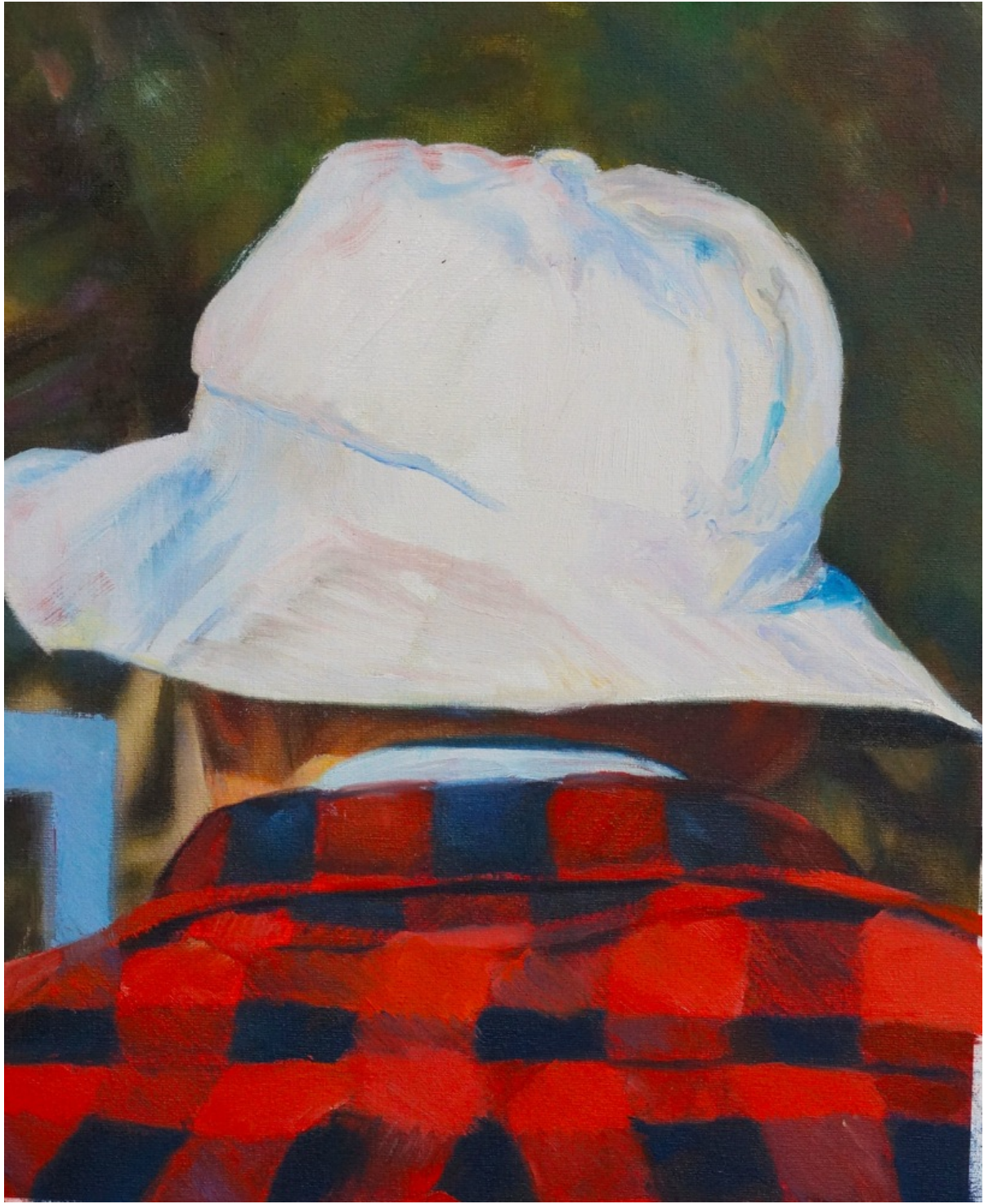